\documentclass[aps,prd,twocolumn]{revtex4}
\usepackage{amsfonts}
\usepackage{amssymb}
\usepackage{amsmath}
\usepackage[usenames,dvipsnames]{xcolor}
\usepackage{float}
\usepackage{accents}
\usepackage{graphicx}
\usepackage{epstopdf}
\usepackage[bookmarks=false,colorlinks=true,pdfstartview=FitV,linkcolor=blue,citecolor=blue,urlcolor=blue,breaklinks=true]{hyperref}
\usepackage{ulem}

\begin{document}

\title{Self-dual solitons in a generalized Chern-Simons baby Skyrme model}
\author{Rodolfo Casana}
\email{rodolfo.casana@gmail.com}
\author{Andr\'{e} C. Santos}
\email{andre$\_$cavs@hotmail.com}
\author{Claudio F. Farias}
\email{cffarias@gmail.com}
\author{Alexsandro L. Mota}
\email{lucenalexster@gmail.com}
\affiliation{Departamento de F\'{\i}sica, Universidade Federal do Maranh\~{a}o,
65080-805, S\~{a}o Lu\'{\i}s, Maranh\~{a}o, Brazil.}

\begin{abstract}
We have shown the existence of self-dual solitons in a type of generalized Chern-Simons baby Skyrme model where the generalized function (depending only in the  Skyrme field) is coupled to the sigma-model term. The consistent implementation of the Bogomol'nyi-Prasad-Sommerfield (BPS) formalism requires the generalizing function becomes the superpotential defining properly the self-dual potential. Thus, we have obtained a topological energy lower-bound (Bogomol'nyi bound) and the self-dual equations satisfied by the fields saturating such a bound. The Bogomol'nyi bound being proportional to the topological charge of the Skyrme field is quantized whereas the total magnetic flux is not. Such as expected in a Chern-Simons model the total magnetic flux and the total electrical charge are proportional to each other.  Thus, by considering the superpotential a well-behaved function in the whole target space we have shown the existence of three types of self-dual solutions: compacton solitons, soliton solutions whose tail decays following an exponential-law $e^{-\alpha r^{2}}$ ($\alpha>0 $), and solitons having a power-law decay $r^{-\beta}$ ($\beta>0$). The profiles of the two last solitons can exhibit a compactonlike behavior. The self-dual equations have been solved numerically and we have depicted the soliton profiles, commenting on the main characteristics exhibited by them.
\end{abstract}

\maketitle

\section{Introduction}

Effective field theories have an important role in physics, especially when
they can provide answers or insights about certain physical properties that
could be difficult or even impossible to be extracted from the respective
underlying higher-energy model. It is the case of the Skyrme model \cite%
{skyrme} proposed to give some critical information about hadronic states,
which result be a hard task when analyzed directly via the Quantum
Chromodynamics (QCD). The proposal of the Skyrme model is to substitute by
means of a scalar triplet the Goldstone bosons produced by the chiral
symmetry breaking \cite{zahed}. Such approach provides an efficient and very
predictive framework for the study of baryon properties \cite{adkins}, as
well as atomic nuclei \cite{houghton}, nuclear matter \cite{ma} and neutron
stars \cite{holt}. The baryons emerge as collective excitations described by
topological solitons called Skyrmions. Some improvements of the Skyrme model
result in very accurate description of baryon masses such as shown in Ref.
\cite{sutcliffe}.

In condensed matter physics, the Skyrmions have achieved a new status in
physics when researchers found promising applications, for example, they
have been studied in systems such as superfluid $^{3}$He \cite{volovik} and
quantum Hall ferromagnets \cite{sarma}. More recently, the discovery of
Skyrmion structures in magnetic materials has been reported, for example,
neutron scattering experiments shown that a Skyrmion crystal was related to
phase transitions in a MnSi bulk \cite{muhlbauer}, Skyrmion behavior was
found in Monte Carlo simulations running on a discretized model of the
chiral magnet in two dimensions \cite{yi}. An important technological step
was made when a Skyrmion phase was obtained on a thin film of the chiral
magnet Fe$_{1-x}$Co$_{x}$Si, which has an energetic stability greater than in
three dimensional systems \cite{yu}. The research on magnetic Skyrmions is a
promising area aiming for technological applications such as data storage
and spintronic.

Recent developments was made on Bose-Einstein condensates \cite{usama} and
chiral nematic liquid crystals \cite{jun}. There are also remarkable works
on superconductivity. Skyrmions has been predicted for K$_{2}$Fe$_{4}$Se$_{5}$
material where superconductivity emerges at room temperature and stable
Skyrmions become Cooper pairs through a quantum anomaly \cite{baskaran}.
Further approaches have been made on this field \cite%
{garaud,garaud2,zyuzin,winyard,vadimov} and also analogies between vortex in
superconductors systems and Skyrmions in magnetic materials. Skyrmion
crystal with a triangular array in magnetic systems was shown to have strong
similarities with Abrikosov vortex lattice in type-II superconductors \cite%
{bogdanov}.

All these planar realizations and residual problems in the Skyrme approach on nuclear physics has inspired the development of a lower dimensional version of the Skyrme model called baby Skyrme model \cite{piette,gisiger1996}. It can be seen as a toy model in $(2+1)$-dimensions that keeps some essential qualitative features of its higher dimensional counterpart.  Among the features explored recently, we point out its topological structure which has enlighten many of its fundamental properties, both qualitative as quantitative, enriching so the range and applicability of the model. Also, gauged versions of the baby Skyrme model have been built by introducing minimal covariant derivatives and the respective dynamical gauge term. Soliton solutions carrying only magnetic flux were obtained in a baby Skyrme model whose gauge field dynamics is governed by Maxwell's term \cite{gladikowski}. Until now it have not been possible to implement the BPS formalism for the baby Skyrme model. However, BPS solitons have been achieved in the gauged nonlinear sigma-model \cite{gisiger1996,schroers}. It leave us to an
important conclusion, the lower-bound of the full baby Skyrme model should
not be below the sigma-model bound \cite{schroers}.

Nevertheless, the so-called restricted baby Skyrme model \cite{gisiger} possess a BPS structure \cite{adam}. For the gauged version with  the Maxwell term, the BPS solutions saturating the energy lower-bound were finally found in Ref. \cite{adam2}. In general, such models have shown itself an interesting avenue of investigations in many issues such as duality between vortices and planar Skyrmions \cite{adam3}, topological phase transitions \cite{adam4}, Bogomol'nyi equation from the strong
necessary conditions \cite{stepien}, gauged BPS baby Skyrmions with
quantized magnetic flux \cite{adam5}, supersymmetry \cite{susy1,susy2,susy3}
and gravitational theories \cite{gravity}.

In $(2+1)$-dimensions besides the Maxwell term with its obvious relevance in
gauged field theories  there is the topological Chern-Simons term which has a central physical role in the emergence of configurations with nonnull total electric charge. The Chern-Simons term plays an important role in field theory \cite{CS1,CS2}, and in the description of some phenomena in bidimensional systems of condensed matter physics, such as fractional statistics \cite{CS3} and the fractional quantum Hall effect \cite{CS4}. In the context of topological defects involving baby Skyrme model, the influence of the Chern-Simons term was studied in Ref. \cite{loginov} obtaining soliton solutions with interesting new features
such as electrical charge; in Ref. \cite{CAdamcs} was analyzed a Lifshitz version of a gauged baby Skyrme model providing BPS solitons; the multi-soliton configurations and the changes with the different potentials was studied in details in Ref. \cite{samoilenka}. Recently, a supersymmetric extension was implemented in Ref. \cite{queiruga}.

The goal of the manuscript is the successful implementation of the BPS
formalism in a generalized version of a gauged baby Skyrme model whose gauge field dynamics is governed solely by Chern-Simons term. Such a model is able to engender BPS compacton and noncompacton solitons, the last ones can exhibit a compactonlike behavior. The manuscript is structured as follows: In Sec. II, we present a Chern-Simons restricted baby Skyrme model where the unsuccessful implementation of the BPS technique has allowed to glimpse the guidelines for the construction of a model able to engender BPS configurations. In Sec. III, based in previous section, we have constructed a true BPS Chern-Simons baby Skyrme model whose successful implementation of the BPS formalism has allowed to obtain a BPS energy-lower bound and the respective self-dual or BPS equations. In Sec. IV we analyze some properties of the rotationally symmetric solitons such as the behavior at boundaries, the magnetic flux and electric charge. The Sec. V is dedicated to the numerical solutions of the BPS equations. Finally, in Sec. VI, we present our conclusions and perspectives.

\section{A non-BPS Chern-Simons restricted baby Skyrme model}

The baby Skyrme model \cite{piette} is a $(2+1)$-dimensional nonlinear field
theory supporting topological solitons described by the Lagrangian density%
\begin{equation}
\mathcal{L}=\frac{\lambda_0^2}{2}\partial _{\mu }\vec{\phi}\cdot \partial
^{\mu }\vec{\phi}-\frac{\lambda ^{2}}{4}(\partial _{\mu }\vec{\phi}\times
\partial _{\nu }\vec{\phi})^{2}-{V}.  \label{lagbsk0}
\end{equation}%
The first contribution stands the sigma-model term, the second one is the
Skyrme term and the third term is the self-interacting potential being, in
principle, a function of the quantity $\hat{n}\cdot \vec{\phi}=\phi _{n}$,
i.e., ${V}\equiv {V}(\phi _{n})$. In the internal space, $\hat{n}$ is an
unitary vector given a preferred direction, the Skyrme field $\vec{\phi}$
defines a triplet of real scalar fields $\vec{\phi}=\left( \phi _{1},\phi
_{2},\phi _{3}\right) $ with fixed norm, $\vec{\phi}\cdot \vec{\phi}=1$,
describing a spherical surface with unitary radius.

In absence of the sigma-model term the resulting one is the so-called restricted baby Skyrme model which is given by
\begin{equation}
\mathcal{L}= -\frac{\lambda ^{2}}{4}(\partial _{\mu }\vec{\phi}\times
\partial _{\nu }\vec{\phi})^{2}-{V}.  \label{lagbsk01}
\end{equation}%

The sigma-model and Skyrme terms are invariants under the global $SO(3)$
symmetry whereas the potential breaks partially it, preserving only the
subgroup $U(1)$ of the target space. The existence of such an unbroken
subgroup $U(1)$ allows to implement a local gauge symmetry by means of the
introduction of a $U(1)$ gauge field whose dynamics, in $(2+1)$-dimensions,
can be governed by the Maxwell action \cite{adam2} or the Chern-Simons
action \cite{loginov,CAdamcs} or both \cite{samoilenka}.

In the remainder of this section we consider a restricted baby Skyrme model
gauged solely with the Chern-Simons term described by the following
Lagrangian density,
\begin{equation}
\mathcal{L}=-\frac{\kappa}{4}\epsilon^{\sigma\mu\nu}A_{\sigma}F_{\mu\nu} -%
\frac{\lambda^{2}}{4} ( D_{\mu}\vec{\phi}\times D_{\nu}\vec{\phi} ) ^{2}-
V(\phi_{n}),  \label{lag0}
\end{equation}
where $\kappa$ is the Chern-Simons coupling constant, $A_{\mu}$ is the
Abelian gauge field and $F_{\mu\nu}=\partial_{\mu}A_{\nu}-\partial_{\nu}A_{%
\mu}$ its the field strength tensor. The minimal covariant derivative of the
Skyrme field $D_{\mu}\vec{\phi}$ is given by
\begin{equation}
D_{\mu}\vec{\phi}=\partial_{\mu}\vec{\phi}+gA_{\mu}\hat{n}\times\vec{\phi}.
\label{covder}
\end{equation}
where $g$ the electromagnetic coupling constant. Here, we consider the gauge
field with mass dimension $1$ and the Skyrme field to be dimensionless.
Hence, both the Chern-Simons coupling constant $\kappa$ and the
electromagnetic one $g$ become dimensionless and, the $\lambda$ coupling
constant has mass dimension $-1/2$.

The gauge field equation obtained from of the Lagrangian density (\ref{lag0}%
) is
\begin{equation}
-\frac{\kappa}{2}\epsilon^{\mu\alpha\beta}F_{\alpha\beta}=J^{\mu},
\label{eq_gauge}
\end{equation}
where $J^{\mu}$ is the conserved gauge current density,
\begin{equation}
J^{\mu}=\lambda^{2}g\left[ \vec{\phi}\cdot( D^{\mu}\vec{\phi}\times
D^{\alpha}\vec{\phi} ) \right] ( \hat{n}\cdot\partial_{\alpha}\vec{\phi} ) .
\label{eq_curr}
\end{equation}
Similarly, the equation of motion of the Skyrme field is
\begin{equation}
\lambda^{2}D_{\mu}\left\{ \left[ \vec{\phi}\cdot( D^{\mu}\vec{\phi}\times
D^{\beta}\vec{\phi} ) \right] D_{\beta}\vec{\phi}\right\} +(\hat{n}\times%
\vec{\phi})\frac{\partial V}{\mathcal{\partial}\phi_{n}} =0.
\label{eq_skyrme}
\end{equation}

We are interested in time-independent solution of the model, thus, we write
down the respective equations of motion. The stationary Gauss law reads
\begin{equation}
\kappa B=\lambda^{2}g^{2}A_{0}(\hat{n}\cdot\partial_{j}\vec{\phi})^{2},
\label{gausslaw}
\end{equation}
where $B=F_{12}=\epsilon_{ij}\partial_{i}A_{j}$ defines the magnetic field.
The stationary Amp{\`{e}}re law is written as
\begin{equation}
\kappa\partial_{j}A_{0}=-\lambda^{2}g( \hat{n}\cdot\partial_{j}\vec{\phi}) Q,
\label{amplaw}
\end{equation}
where we have introduced the quantity $Q$ defined by
\begin{equation}
Q=\vec{\phi}\cdot( D_{1}\vec{\phi}\times D_{2}\vec{\phi}) .  \label{Q1}
\end{equation}
The respective equation of motion of the Skyrme field is
\begin{eqnarray}
0 & =&(\hat{n}\times\vec{\phi}) \frac{\partial V}{\mathcal{\partial}\phi_{n}}%
+\lambda^{2}\epsilon_{ij}D_{i} ( Q D_{j}\vec{\phi} )  \notag \\[0.2cm]
&& +\lambda^{2}g^{2}(\hat{n}\times\vec{\phi}) \partial_{j}\left[ \left(
A_{0}\right) ^{2}(\hat{n}\cdot\partial_{j}\vec{\phi}) \right] .  \label{ss1}
\end{eqnarray}

\subsection{BPS formalism: The frustrated implementation}

In the stationary regime, the energy density corresponding to the model (\ref%
{lag0}) reads
\begin{equation}
\varepsilon=\frac{\lambda^{2}}{2}g^{2}\left( A_{0}\right) ^{2}( \hat{n}%
\cdot\partial_{j}\vec{\phi}) ^{2}+\frac{\lambda^{2}}{2}Q^{2}+V,  \label{eex1}
\end{equation}
where we have used the identity
\begin{equation}
( D_{i}\vec{\phi}\times D_{j}\vec{\phi}) ^{2}=2Q^{2}.
\end{equation}
We first use the Gauss law (\ref{gausslaw}), to express $A_{0}$ in terms of
the magnetic field, such that the energy density (\ref{eex1}) becomes
\begin{equation}
\varepsilon=\frac{1}{2}\frac{\kappa^{2}}{\lambda^{2}g^{2}}\frac{B^{2}}{(
\hat{n} \cdot\partial_{j}\vec{\phi} ) ^{2}}+\frac{\lambda^{2}}{2}Q^{2}+V.
\label{enx1}
\end{equation}
In order to perform the implementation of the BPS formalism we introduce
into (\ref{enx1}) two functions $W(\phi_{n})$ and $Z(\phi_{n})$ to be
determined \textit{a posteriori}. Thus, after some algebraic manipulations
the energy density (\ref{enx1}) can be rewritten as
\begin{eqnarray}
\varepsilon & =&\frac{1}{2}\frac{\kappa^{2}}{\lambda^{2}g^{2}}\frac{\left[
B\pm( \hat{n}\cdot\partial_{j}\vec{\phi}) ^{2}W\right] ^{2}}{( \hat{n}%
\cdot\partial_{j}\vec{\phi}) ^{2}}+\frac{\lambda^{2}}{2}\left( Q\mp Z\right)
^{2}  \notag \\[0.2cm]
&& \mp\frac{\kappa^{2}}{\lambda^{2}g^{2}}BW\pm\lambda^{2}QZ  \notag \\[0.2cm]
&& +V-\frac{\lambda^{2}}{2}Z^{2}-\frac{1}{2}\frac{\kappa^{2}}{%
\lambda^{2}g^{2}}( \hat{n}\cdot\partial_{j}\vec{\phi}) ^{2}W^{2}.
\label{denE01}
\end{eqnarray}
This procedure is already utilized in literature with the aim to attain a
successfully implementation of the BPS formalism. For example, it was used in the case of Skyrmions \cite{adam2,CAdamcs} and some generalized versions of Maxwell-Higgs model \cite{casanavts}.

By using the relation
\begin{equation}
Q=\vec{\phi}\cdot( \partial_{1}\vec{\phi}\times\partial_{2}\vec{\phi})
+g\epsilon_{ij}A_{i}( \hat{n}\cdot\partial_{j}\vec{\phi}) ,
\end{equation}
and expressing the magnetic field as $B=-\epsilon_{ij}\partial_{j}A_{i}$,
the energy density (\ref{denE01}) becomes
\begin{eqnarray}
\varepsilon & =&\frac{1}{2}\frac{\kappa^{2}}{\lambda^{2}g^{2}}\frac{\left[
B\pm( \hat{n}\cdot\partial_{j}\vec{\phi}) ^{2}W\right] ^{2}}{( \hat{n}%
\cdot\partial_{j}\vec{\phi}) ^{2}}+\frac{\lambda^{2}}{2}\left( Q\mp Z\right)
^{2}  \notag \\[0.2cm]
&& \pm\lambda^{2}Z\vec{\phi}\cdot( \partial_{1}\vec{\phi}\times\partial _{2}%
\vec{\phi} )  \notag \\[0.2cm]
&& \pm\epsilon_{ij}\left[ ( \partial_{j}A_{i} ) \frac{\kappa^{2}}{\lambda
^{2}g^{2}}W+A_{i}\lambda^{2}gZ ( \hat{n}\cdot\partial_{j}\vec{\phi} ) \right]
\notag \\[0.2cm]
&& +V-\frac{\lambda^{2}}{2}Z^{2}-\frac{1}{2}\frac{\kappa^{2}}{%
\lambda^{2}g^{2}}( \hat{n}\cdot\partial_{j}\vec{\phi}) ^{2}W^{2}.
\label{enovac1}
\end{eqnarray}
The term $\vec{\phi}\cdot( \partial_{1}\vec{\phi}\times\partial_{2}\vec{\phi}%
)$ in the second row is related to the topological degree (topological
charge or winding number) of the Skyrme field which is defined by
\begin{equation}
\mathrm{deg}[\vec{\phi}]=-\frac{1}{4\pi}\!\int d^{2}\mathbf{x}\;\vec{\phi}%
\cdot( \partial_{1}\vec{\phi}\times\partial_{2}\vec{\phi} ) =k.
\label{chargsk}
\end{equation}
where $k$ is a non-null integer.

The implementation of the BPS formalism will be completed in Eq. (\ref%
{enovac1}) if we transform the third row in a total derivative and set to be
null the fourth row. Thus, the first goal is attained by establishing the
following relation
\begin{equation}
\frac{\kappa^{2}}{\lambda^{2}g^{2}}\partial_{j}W=\lambda^{2}gZ ( \hat{n}%
\cdot\partial_{j}\vec{\phi}) ,
\end{equation}
which allows to determine the function $Z$ in terms of $W$,
\begin{equation}
Z=\frac{\kappa^{2}}{\lambda^{4}g^{3}}\frac{\partial W}{\partial\phi_{n}}.
\end{equation}
Our second objective provides the potential
\begin{equation}
V=\frac{1}{2}\frac{\kappa^{4}}{\lambda^{6}g^{6}}\left( \frac{\partial W}{%
\partial\phi_{n}}\right) ^{2}+\frac{1}{2}\frac{\kappa^{2}}{\lambda ^{2}g^{2}}
( \hat{n}\cdot\partial_{j}\vec{\phi} ) ^{2}W^{2},  \label{potftdo}
\end{equation}
where we see the function $W(\phi_{n})$ plays the role of a
``superpotential" such as it has been pointed in literature \cite%
{adam2,CAdamcs}.

By implementing all that, the energy density becomes
\begin{eqnarray}
\varepsilon & =&\frac{\kappa^{2}}{2\lambda^{2}g^{2}}\!\frac{\left[ B\pm (%
\hat{n}\cdot\partial_{j}\vec{\phi})^{2}W\right] ^{2}}{\left( \hat{n}%
\cdot\partial_{j}\vec{\phi}\right) ^{2}}+\frac{\lambda^{2}}{2}\left[ Q\mp%
\frac{\kappa^{2}}{\lambda^{4}g^{3}}\frac{\partial W}{\partial\phi_{n}}\right]
^{2}  \notag \\[0.2cm]
&& \hspace{-0.25cm}\pm\frac{\kappa^{2}}{\lambda^{2}g^{3}}\frac{\partial W}{%
\partial\phi_{n}}\vec{\phi}\cdot( \partial_{1}\vec{\phi}\times\partial _{2}%
\vec{\phi} ) \pm\frac{\kappa^{2}}{\lambda^{2}g^{2}}\epsilon_{ij}\partial_{j}
( WA_{i} ) .
\end{eqnarray}
The terms in the squared brackets would be the BPS equations, the third term related to the topological charge of the  Skyrme field provides the BPS limit for the total energy and the fourth term being a total derivative would give null contribution to the total energy if $\lim_{\phi_n\rightarrow 1}W(\phi_n)= \lim_{ \mathbf{x}
\rightarrow \infty} W(\phi_n)=0$.

Until here the implementation of the BPS formalism looks like successful, however, there is a contradiction with the hypothesis on the functional dependence of the potential introduced in the Lagrangian density (\ref{lag0}) because now the BPS potential (\ref{potftdo}) is not a solely function of the Skyrme field due to it also contains its derivative. Consequently, the stationary Euler-Lagrange equation (\ref{ss1}) of the Skyrme fields is not recovered from such BPS equations.

Notwithstanding our first attempt to implement the BPS formalism was
unsuccessful, the potential (\ref{potftdo}) indicates a way how to introduce
new terms in the model (\ref{lag0}) with the aim to turn it in a one capable
to engender BPS configurations. The modified model with such a property is
introduced in next section.

\section{A BPS Chern-Simons baby Skyrme model}

The previous procedure suggests the existence of BPS configurations can be well established in a modified version of the model (\ref{lag0}).  Such a modification is not arbitrary,  the guidelines to perform such a change is given by the derivative term of Eq. (\ref{potftdo}) which indicate us that a term proportional to $(\hat{n}\cdot D_{\mu}\vec{\phi}) ^{2} W^{2}$ must be introduced in the Lagrangian density (\ref{lag0}). Thus, the new model capable to engender BPS configurations is described by the following Lagrangian density
\begin{eqnarray}
\mathcal{L} & =&-\frac{\kappa}{4}\epsilon^{\sigma\mu\nu}A_{\sigma}F_{\mu\nu
}-\frac{\lambda^{2}}{4} ( D_{\mu}\vec{\phi}\times D_{\nu}\vec{\phi} ) ^{2}
\notag \\[0.2cm]
& &+\frac{1}{2}\frac{\kappa^{2}}{\lambda^{2}g^{2}}(\hat{n}\cdot D_{\mu}\vec{%
\phi}) ^{2} W^{2}- \mathcal{V},  \label{lag1}
\end{eqnarray}
where both the dimensionless function $W$ and the potential $\mathcal{V}$
depend only in the variable $\phi_{n}$. The third term modifies the dynamics
of the component along the direction $\hat{n}$ of the Skyrme field. In this
way, the last two terms break partially the $SO(3)$ symmetry preserving the $%
U(1)$ subgroup of this symmetry.

The term $(\hat{n}\cdot D_{\mu }\vec{\phi})^{2}$ in (\ref{lag1}) can be expressed in the following form
\begin{equation}
(\hat{n}\cdot D_{\mu }\vec{\phi})^{2}=D_{\mu }\vec{\phi}\cdot D^{\mu }\vec{%
\phi}-(\hat{n}\times D_{\mu }\vec{\phi})^{2},
\end{equation}%
allowing to express the Lagrangian density (\ref{lag1}) as
\begin{eqnarray}
\mathcal{L} &=&-\frac{\kappa }{4}\epsilon ^{\sigma \mu \nu }A_{\sigma
}F_{\mu \nu }+\frac{1}{2}\frac{\kappa ^{2}}{\lambda ^{2}g^{2}}W^{2}D_{\mu }
\vec{\phi}\cdot D^{\mu }\vec{\phi}  \notag \\[0.2cm]
&&\hspace{-0.5cm} -\frac{\lambda ^{2}}{4}(D_{\mu }\vec{\phi}\times D_{\nu }%
\vec{\phi})^{2}- \frac{1}{2}\frac{\kappa ^{2}}{\lambda ^{2}g^{2}}(\hat{n}%
\times D_{\mu }\vec{ \phi})^{2}W^{2}-\mathcal{V}.\quad\quad\label{lag1a}
\end{eqnarray}
The second term is a generalized gauged sigma-model with the function $W$ playing the role of the generalizing function, the third one is the gauged Skyrme term. In other words, the new model (\ref{lag1}) is a type of generalized Chern-Simons baby Skyrme model modified by the term proportional to $(\hat{n}\times D_{\mu}\vec{ \phi})^{2} W^{2}$.

We point out the gauge field equation coming from the Lagrangian density (%
\ref{lag1}) is exactly the same given by Eq. (\ref{eq_gauge}), i.e., the
introduction of the function ${W}(\phi_{n})$ does not modify the gauge field
equation of motion.

The equation of motion of the Skyrme field obtained from the Lagrangian
density (\ref{lag1}) is
\begin{align}
0 & =\lambda^{2}D_{\mu}\left\{ (D_{\beta}\vec{\phi})\left[ \vec{\phi}%
\cdot(D^{\mu}\vec{\phi}\times D^{\beta}\vec{\phi}) \right] \right\}  \notag
\\[0.2cm]
& +(\hat{n}\times\vec{\phi}) \left\{ \frac{\partial\mathcal{V} }{\mathcal{%
\partial}\phi_{n}}+\frac{\kappa^{2}}{\lambda^{2}g^{2}} \partial_{\mu}\left[
( \hat{n}\cdot\partial^{\mu}\vec{\phi}) W^{2}\right] \right.  \notag \\%
[0.2cm]
& \hspace{1.6cm}\left. -\frac{\kappa^{2}}{2\lambda^{2}g^{2}} ( \hat{n}%
\cdot\partial_{\nu}\vec{\phi} ) ^{2}\frac{\partial W^{2}}{\mathcal{\partial }%
\phi_{n}}\right\} ,
\end{align}
whose stationary version reads,
\begin{align}
0 & =\lambda^{2}\epsilon_{ij}D_{i}\left( Q D_{j}\vec{\phi}\right) +(\hat{n}%
\times\vec{\phi}) \frac{\partial\mathcal{V}}{\mathcal{\partial}\phi_{n}}
\notag  \label{feq} \\[0.2cm]
& +\lambda^{2}g^{2}(\hat{n}\times\vec{\phi}) \partial_{j} \left[ \left(
A_{0}\right) ^{2} ( \hat{n}\cdot\partial_{j}\vec{\phi}) \right]  \notag \\%
[0.2cm]
& -\frac{\kappa^{2}}{\lambda^{2}g^{2}}( \hat{n}\times\vec{\phi}) \partial
_{j}\left[ ( \hat{n}\cdot\partial_{j}\vec{\phi}) W^{2}\right]  \notag \\%
[0.2cm]
& +\frac{\kappa^{2}}{2\lambda^{2}g^{2}}( \hat{n}\times\vec{\phi}) ( \hat {n}%
\cdot\partial_{j}\vec{\phi}) ^{2}\frac{\partial W^{2}}{\mathcal{\partial }%
\phi_{n}}.
\end{align}

In the following, we going to show how to build the BPS formalism
determining the self-interacting potential $\mathcal{V}(\phi_{n})$ which
allows to obtain a lower-bound for the energy and the self-dual equations
satisfied by the soliton configurations saturating a BPS bound.

\subsection{The BPS configurations}

The stationary energy density is given by
\begin{eqnarray}
\varepsilon & =&\frac{\lambda^{2}}{2}g^{2}\left( A_{0}\right) ^{2} (\hat {n}%
\cdot\partial_{j}\vec{\phi})^{2}+\frac{\lambda^{2}}{2}Q^{2}  \notag \\[0.2cm]
&& +\frac{\kappa^{2}}{2\lambda^{2}g^{2}}(\hat{n}\cdot\partial_{j}\vec{\phi }%
)^{2} W^{2}+\mathcal{V}.  \label{energy0}
\end{eqnarray}
By using the Gauss law (\ref{gausslaw}), the energy density reads
\begin{eqnarray}
\varepsilon & =&\frac{1}{2}\frac{\kappa^{2}}{\lambda^{2}g^{2}}\frac{B^{2}}{(
\hat{n}\cdot\partial_{j}\vec{\phi}) ^{2}}+\frac{1}{2}\frac{\kappa^{2}}{%
\lambda^{2}g^{2}}( \hat{n}\cdot\partial_{j}\vec{\phi}) ^{2}W^{2}  \notag \\%
[0.2cm]
&& +\frac{\lambda^{2}}{2}Q^{2}+\mathcal{V}.  \label{energy}
\end{eqnarray}
After some algebraic manipulations, we write the energy density (\ref{energy}%
) as
\begin{eqnarray}
\varepsilon & =&\frac{1}{2}\frac{\kappa^{2}}{\lambda^{2}g^{2}}\frac{\left[
B\pm( \hat{n}\cdot\partial_{j}\vec{\phi}) ^{2}W\right] ^{2}}{\left( \hat {n}%
\cdot\partial_{j}\vec{\phi}\right) ^{2}}+\frac{\lambda^{2}}{2}\left( Q\mp%
\frac{\sqrt{2\mathcal{V}}}{\lambda}\right) ^{2}  \notag \\[0.2cm]
&& \hspace{-0.5cm}\pm\lambda\vec{\phi}\cdot(\partial_{1}\vec{\phi}%
\times\partial_{2}\vec{\phi}) \sqrt{2\mathcal{V}}  \notag \\[0.3cm]
& & \hspace{-0.5cm}\mp\frac{\kappa^{2}}{\lambda^{2}g^{2}}\epsilon_{ij}\left[
\left( \partial_{i}A_{j}\right) W+A_{j}\frac{\lambda^{3}g^{3}}{\kappa^{2}}(
\hat{n}\cdot\partial_{i}\vec{\phi}) \sqrt{2\mathcal{V}}\right] \!. \quad\;\;
\;  \label{energy2}
\end{eqnarray}
The term in the second row is related to the topological charge of the
Skyrme field and as we will see below it provides the Bogomol'nyi limit for
the total energy. At this point, the implementation of the BPS formalism
will be completed if we transform the terms of the third row in Eq. (\ref%
{energy2}) in a total derivative. This is achieved by requiring the function
$W(\phi_{n})$ satisfies the following constraint
\begin{equation}
\partial_{i}W=\frac{\partial W}{\partial\phi_{n}} ( \hat{n}\cdot\partial_{i}
\vec{\phi} ) =\frac{\lambda^{3}g^{3}}{\kappa^{2}} \sqrt{2\mathcal{V}}(\hat {n%
}\cdot\partial_{i}\vec{\phi} ),  \label{condition0}
\end{equation}
which allows us determine the self-interacting potential able to engender
self-dual configurations,
\begin{equation}
\mathcal{V}(\phi_{n})=\frac{\kappa^{4}}{2\lambda^{6}g^{6}}\left( \frac{%
\partial W}{\partial\phi_{n}}\right) ^{2}.  \label{condition}
\end{equation}
Such a equation shows clearly the role of a superpotential played by the
function $W(\phi_{n})$ into the model (\ref{lag1}).

The Eqs. (\ref{condition0}) and (\ref{condition}) allow to write the energy density (\ref{energy2}) as
\begin{eqnarray}
\varepsilon & =&\!\frac{\kappa^{2}}{2\lambda^{2}g^{2}}\frac{\left[ B\pm(
\hat{n}\cdot\partial_{j}\vec{\phi}) ^{2}W\right] ^{2}}{\left( \hat{n}%
\cdot\partial_{j}\vec{\phi}\right) ^{2}}+\frac{\lambda^{2}}{2}\!\left( \!Q\mp%
\frac{\kappa^{2}}{\lambda^{4}g^{3}}\frac{\partial W}{\partial\phi_{n}}%
\!\right) ^{2}  \notag \\[0.2cm]
& & \hspace{-0.5cm}\pm\frac{\kappa^{2}}{\lambda^{2}g^{3}}\vec{\phi}\cdot(
\partial_{1}\vec{\phi}\times\partial_{2}\vec{\phi} ) \frac{\partial W}{%
\partial\phi_{n}}\mp\frac{\kappa^{2}}{\lambda^{2}g^{2}}\epsilon
_{ij}\partial_{i}(WA_{j}).  \label{energy2a}
\end{eqnarray}

The superpotential $W(\phi_{n})$ must be constructed or proposed in order to the potential (\ref{condition}) can satisfy the vacuum condition $\mathcal{V} (\phi_{n}\rightarrow1) \rightarrow0$, and eliminate the contribution of the total derivative, $\epsilon_{ij}\partial_{i}\left( WA_{j}\right)$, in (\ref{energy2a}) to the total energy, i.e.,
\begin{equation}
\int d^{2}\mathbf{x}\,\varepsilon_{ij}\partial_{i}(A_{j}W)=0.
\end{equation}
Thus, we consider superpotential satisfying the following boundary
conditions
\begin{equation}
\lim_{\phi_{n}\rightarrow1}\frac{\partial W}{\partial\phi_{n}} =\lim _{|%
\mathbf{x}|\rightarrow\infty}\frac{\partial W}{\partial\phi_{n}} =0,
\label{cw2}
\end{equation}
\begin{equation}
\lim_{\phi_{n}\rightarrow1}W(\phi_{n})=\lim_{|\mathbf{x}|\rightarrow\infty}
W(\phi_{n}) =0,  \label{cw1}
\end{equation}
respectively. Consequently, we write the total energy as
\begin{equation}
\mathcal{E}=\int d^{2}\mathbf{x}~\varepsilon=\mathcal{E}_{BPS}+\check {%
\mathcal{E}},  \label{energy2b}
\end{equation}
where $\mathcal{E}_{BPS}$ defining the energy lower-bound is
\begin{equation}
\mathcal{E}_{BPS}=\pm\frac{\kappa^{2}}{\lambda^{2}g^{3}}\int d^{2}\mathbf{x}~%
\vec{\phi}\cdot( \partial_{1}\vec{\phi}\times\partial_{2}\vec{\phi} ) \frac{%
\partial W}{\partial\phi_{n}},  \label{bpsenergy}
\end{equation}
and $\check{\mathcal{E}}$ is given by
\begin{eqnarray}
\check{\mathcal{E}} & =&\int d^{2}\mathbf{x}\left\{ \frac{1}{2}\frac {%
\kappa^{2}}{\lambda^{2}g^{2}}\frac{\left[ B\pm( \hat{n}\cdot\partial_{j}
\vec{\phi} )^{2}W\right] ^{2}}{ ( \hat{n}\cdot\partial_{j}\vec{\phi} ) ^{2}}%
\right.  \notag \\[0.25cm]
& & \hspace{1.25cm}\left. +\frac{\lambda^{2}}{2}\left( Q\mp\frac{\kappa^{2}}{%
\lambda^{4}g^{3}}\frac{\partial W}{\partial\phi_{n}}\right) ^{2}\right\} .
\end{eqnarray}

From the expression of the total energy (\ref{energy2b}) we observe the
following inequality is always satisfied%
\begin{equation}
\mathcal{E}\geq\mathcal{E}_{BPS}.  \label{ineq}
\end{equation}
The lower-bound is saturated, i.e., $\check{\mathcal{E}}=0$, if the fields
satisfy\ the self-dual or BPS equations%
\begin{eqnarray}
Q & =& \pm\frac{\kappa^{2}}{\lambda^{4}g^{3}}\frac{\partial W}{\partial
\phi_{n}},  \label{bps1} \\[0.2cm]
B & =& \mp\left( \hat{n}\cdot\partial_{j}\vec{\phi}\right) ^{2}W.
\label{bps2}
\end{eqnarray}
These BPS configurations can be considered as classical solutions related to an extended supersymmetric theory \cite{witten} of the model (\ref{lag1}).

After a long algebraic work, it can be shown starting from the BPS equations we recover the stationary Euler-Lagrange equations provided by the Lagrangian density (\ref{lag1}) which are given by the Gauss law (\ref{gausslaw}), the Amp\`ere law (\ref{amplaw}), and the Skyrme field equation (\ref{feq}).

\section{Rotationally symmetric BPS Skyrmions}

Without loss of generality, we set $\hat{n}\equiv\hat{n}_{3}=\left(0,0,1\right)$, such that $\hat{n}\cdot\vec{\phi}=\phi_{n}=\phi_{3}$, and consider the following  \textit{Ansatz} for the Skyrme field,
\begin{equation}
\left(
\begin{array}{c}
\phi_{1} \\
\phi_{2} \\
\phi_{3}%
\end{array}
\right) =\left(
\begin{array}{c}
\sin f(r)\cos N\theta \\
\sin f(r)\sin N\theta \\
\cos f(r)%
\end{array}
\right) ,  \label{ansatzf}
\end{equation}
where $N=\mathrm{deg}[\vec{\phi}]$ is the winding number of the Skyrme field. For the gauge field, we use
\begin{equation}
A_{k}=-\varepsilon_{kj}\frac{x_{j}}{gr^{2}}[a(r)-N],
\label{ansatza}
\end{equation}
thus, the magnetic field is given by
\begin{equation}
B=\frac{1}{gr}\frac{da}{dr}.
\end{equation}

The functions $f(r)$ and $a(r)$ are well behaved and must satisfy the
boundary conditions:
\begin{equation}
f(0)=\pi,~\ \ f(\infty)=0,
\end{equation}
\begin{equation}
a(0)=N,~\ a(\infty)=a_{\infty},  \label{bcc1}
\end{equation}
where $a_{\infty}$ is a finite quantity.

In order to perform our analysis we introduce the following field
redefinition
\begin{equation}
h=\frac{1}{2} ( 1+\phi_{3} ) =\frac{1}{2} ( 1+\cos f ) ,  \label{trans}
\end{equation}
with the field $h(r)$ satisfying the boundary conditions%
\begin{equation}
h(0)=0,\ \ h(\infty)=1.  \label{bcc2}
\end{equation}

We consider superpotentials $W$ satisfying the following boundary conditions
at origin
\begin{equation}
\lim_{r\rightarrow 0}W(h)=W_{0},\;\lim_{r\rightarrow 0}\frac{dW}{dr}=cte,
\end{equation}%
($W_{0}$ a finite quantity). From equations  (\ref{cw2}) and (\ref{cw1}) we obtain the boundary conditions for $r\rightarrow\infty$,
\begin{equation}
\lim_{r\rightarrow \infty }W(h)=0,\;\lim_{r\rightarrow \infty }\frac{dW}{dr}%
=0,
\end{equation}
the last one guarantees the superpotential be able to generate a potential satisfying the vacuum condition,
\begin{equation}
\mathcal{V}(\infty )\equiv \mathcal{V}(h=1)=0.
\end{equation}

Under the \textit{Ansatz}, the BPS equations become
\begin{eqnarray}
\frac{1}{r}\frac{da}{dr} &=&\mp4g\left( \frac{dh}{dr}\right) ^{2}W,
\label{bps1h} \\[0.2cm]
\frac{a}{r}\frac{dh}{dr} &=&\mp\frac{\kappa^{2}}{4\lambda^{4}g^{3}}
\frac{\partial W}{\partial h}  .  \label{bps2h}
\end{eqnarray}
Similarly, the BPS energy density reads
\begin{equation}
\varepsilon_{_{BPS}}=\frac{4\kappa^{2}}{\lambda^{2}g^{2}}W^{2}\left( \frac{dh%
}{dr}\right) ^{2}+\frac{\kappa^{4}}{4\lambda^{6}g^{6}}\left( \frac{\partial W%
}{\partial h}\right) ^{2},  \label{edbps}
\end{equation}
while the BPS energy (\ref{bpsenergy}) becomes
\begin{equation}
\mathcal{E}_{BPS}=\pm2\pi N\frac{\kappa^{2}}{\lambda^{2}g^{3}}W\left( 0\right) .
\end{equation}

The total magnetic flux $\Phi$ is computed to be
\begin{equation}
\Phi=2\pi\int_{0}^{\infty}rdr\,B=\frac{2\pi}{g}\left[ a_{\infty}-N\right] ,
\label{flux}
\end{equation}
which is, in general, a nonquantized quantity.

By integrating the Gauss law (\ref{gausslaw}), we also obtain the total
electric charge $\mathcal{Q}_{\text{em}}$ being proportional to the total
magnetic flux $\Phi$,
\begin{equation}
\mathcal{Q}_{\text{em}}=\frac{\kappa}{g} \Phi,  \label{electricmagnetic}
\end{equation}
where the total electric charge was defined by
\begin{equation}
\mathcal{Q}_{\text{em}}=g\lambda^{2}\int d^{2}\mathbf{x}\,A_{0} ( \hat{n}%
\cdot\partial_{j}\vec{\phi} ) ^{2}.
\end{equation}
In Sec. V, the numerical analysis has shown, for sufficiently large values
of $g$, the magnetic flux becomes almost a topologically quantized
observable. This effective quantization implies the total electric charge is
also quantized.

\subsection{Behavior of the profiles at origin}

We first solve the BPS equations (\ref{bps1h}) and (\ref{bps2h}) around $r=0$
by considering the boundary conditions%
\begin{equation}
h(0)=0,~a(0)=N,~W(0)=W_{0}.
\end{equation}
The superpotential $W(h) $ is considered to be a well-behaved function  such that behavior for the field profiles $h(r)$ and $a(r)$ result
\begin{eqnarray}
h(r) &=&-\frac{\kappa^{2}(W_{h}) _{h=0}}{8N\lambda
^{4}g^{3}}r^{2}  \notag \\[0.2cm]
& & +\frac{\kappa^{4}(W_{h}) _{h=0}\left( W_{hh}\right) _{h=0}}{%
128N^{2}\lambda^{8}g^{6}}r^{4}+..., \\[0.3cm]
a(r) &=& N-\frac{\kappa^{4}W_{0}(W_{h}) _{h=0}^{2}}{16N^{2}\lambda
^{8}g^{5}}r^{4}  \notag \\[0.2cm]
&& +\frac{\kappa^{6}W_{0}(W_{h}) _{h=0}^{2}\left( W_{hh}\right) _{h=0}}{%
96N^{3}\lambda^{12}g^{8}}r^{6}+...,
\end{eqnarray}
where $W_{h}$ and $W_{hh}$ represent the first and second derivatives of $%
W(h)$ with respect to $h$, respectively.

The magnetic field behavior near to the origin is
\begin{eqnarray}
\left\vert B(r)\right\vert &=& \frac{\kappa^{4}W_{0}(W_{h})_{h=0}^{2}%
}{4N^{2}\lambda^{8}g^{6}}r^{2}  \notag  \label{magr=0} \\[0.2cm]
& & -\frac{\kappa^{6}W_{0}(W_{h})_{h=0}^{2}(W_{hh})_{h=0}}{16N^{3}\lambda
^{12}g^{9}}r^{4}+...,
\end{eqnarray}
and the BPS energy density behaves as%
\begin{eqnarray}
\varepsilon_{_{BPS}} &=& \frac{\kappa^{4}(W_{h})_{h=0}^{2}}{%
4\lambda^{6}g^{6}}  \notag \\[0.2cm]
&& \hspace{-1cm}+\frac{\kappa^{6}(W_{0})^{2}(W_{h})_{h=0}^{2}} {%
4N^{2}\lambda^{10}g^{9}}\left[ g-\frac{N(W_{hh})_{h=0}}{4(W_{0})^{2}}\right]
r^{2} +....\quad  \label{ebpsr=0}
\end{eqnarray}
For small values of $g$, the BPS energy density has greater amplitudes.

\subsection{Behavior of the profiles for large values of $r$}

The analysis for sufficiently large values of $r$ is performed by
considering the boundary conditions%
\begin{equation}
h(R)=1,~a(R)=a_{R},~W(R)=0,
\end{equation}
where $R>0$ and $a_{R}$ a real number. If $R$ is finite, it defines the
maximum radius (size) of a topological defect named compacton. On the other
hand, if $R\rightarrow\infty$, we have a topological defect whose tail
decays following or a exponential law or a power-law.

We have considered the superpotential $W(h)$ behaves when $r\rightarrow R$
as
\begin{equation}
W (h) \approx(1-h) ^{\sigma},
\end{equation}
with the parameter $\sigma>1$. Until now we have found three types of
soliton solutions:
\begin{itemize}
\item[(i)] For $1<\sigma<2$, there are compacton solitons;
\item[(ii)] for $\sigma=2$, the soliton tail decays following a exponential law type $e^{-\alpha r^{2}},\,\alpha>0$; and
\item[(iii)] for $\sigma>2$, the solitons have a power-law decay type $r^{-\beta},\,\beta>0$.
\end{itemize}

\subsubsection{Behavior of the profiles for $1<\protect\sigma<2$}

We consider the compacton has a maximum radius $R$ and the superpotential
behaves at $r=R$ as
\begin{equation}
W(h)\approx W_{R}(1-h)^{\sigma}.
\end{equation}
By considering it and solving the BPS equations we found the profile
functions behave as%
\begin{eqnarray}
h(r) &=& 1-\left( C_{R}^{(h) }\right) ^{1/(2-\sigma
)}(R-r)^{1/(2-\sigma)}+..., \\[0.2cm]
a(r) &=& a_{R}+C_{R}^{\left( a\right) }(R-r)^{2\sigma/(2-\sigma
)}+...,
\end{eqnarray}
where%
\begin{eqnarray}
C_{R}^{(h) } &=&\frac{\sigma(2-\sigma)\kappa^{2}RW_{R}}{%
4\lambda^{4}g^{3}a_{R}}, \\[0.2cm]
C_{R}^{ ( a ) } &=&\frac{2gRW_{R}}{\sigma(2-\sigma)}\left( C_{R}^{(h)
}\right) ^{(2+\sigma)/(2-\sigma)}.
\end{eqnarray}

\subsubsection{Behavior of the profiles for $\protect\sigma=2$}

For the superpotential whose behavior for $r\rightarrow \infty$ is given by
\begin{equation}
W(h)\approx W_{\infty}^{(2)}(1-h)^{2},
\end{equation}
the profile functions behave as%
\begin{eqnarray}
h(r) &=& 1-Ce^{-M^{2}r^{2}}+..., \\[0.2cm]
a(r) &=& a_{\infty}+\,2gW_{\infty}^{(2)}C^{4}M^{2}{r}%
^{2}e^{-4M^{2}r^{2}}+...,
\end{eqnarray}
where the quantity $M$ is given by
\begin{equation}
M^{2}=\frac{\kappa^{2}W_{\infty}^{(2)}}{4\lambda^{4}g^{3}a_{\infty}}.
\label{massa2}
\end{equation}
It verifies the soliton tail has an exponential decay law.

\subsubsection{Behavior of the profiles for {$\protect\sigma>2$}}

We consider the superpotential for $r\rightarrow\infty$ behave as
\begin{equation}
W(h)\approx W_{\infty}(1-h)^{\sigma},
\end{equation}
the profiles has the following behavior%
\begin{eqnarray}
h(r) &=& 1-\left( \frac{C^{\left( \infty\right) }}{r^{2}}\right)
^{1/(\sigma-2)}+..., \\[0.2cm]
a(r) &=& a_{\infty}+\frac{8gW_{\infty}}{\left( \sigma^{2}-4\right) }%
\left( \frac{C^{\left( \infty\right) }}{r^{2}}\right) ^{(2+\sigma
)/(\sigma-2)}+..., \quad
\end{eqnarray}
where%
\begin{equation}
C^{\left( \infty\right) }=\frac{8\lambda^{4}g^{3}a_{\infty}}{\kappa
^{2}W_{\infty}\sigma\left( \sigma-2\right) }.
\end{equation}
We see the profiles follows a power-law decay but the gauge field decays
more fast than the Skyrme field.

\section{Numerical solution of the BPS equations}

\subsection{Compacton solutions}

We have solved the BPS equations (\ref{bps1h}) and (\ref{bps2h}) for the
following superpotential
\begin{equation}
W(h)=W_{0}(1-h)^{3/2},
\end{equation}
which provides the potential
\begin{equation}
\mathcal{V}(h)=\frac{9\kappa^{4}W_{0}^{2}}{32\lambda^{6}g^{6}}(1-h).
\end{equation}
It is the equivalent to the well-known \textquotedblleft old baby Skyrme
potential" \cite{adam2}.

\begin{figure}[H]
\centering{\includegraphics[height=6.5cm]{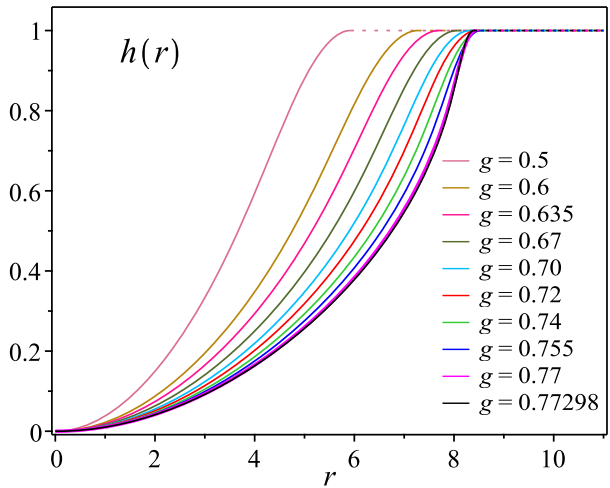}\vspace{-0.25cm} }
\caption{The Skyrme field profile $h(r)$. }
\label{fig1}
\end{figure}

\begin{figure}[H]
\centering{\includegraphics[height=6.5cm]{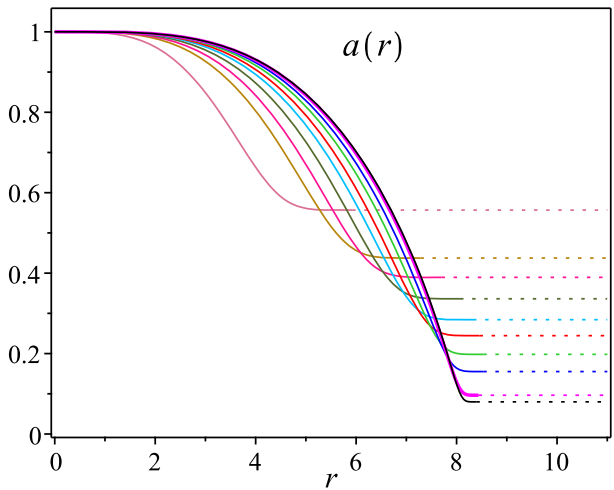}\vspace{-0.25cm} }
\caption{The gauge field profile $a(r)$.}
\label{fig2}
\end{figure}

\begin{figure}[H]
\centering{\includegraphics[height=6.5cm]{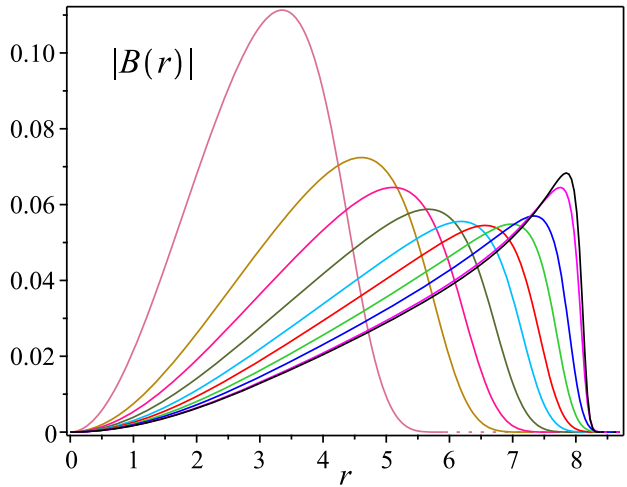}\vspace{-0.25cm} }
\caption{The magnetic field $B(r)$. }
\label{fig3}
\end{figure}

\begin{figure}[H]
\centering{\includegraphics[height=6.5cm]{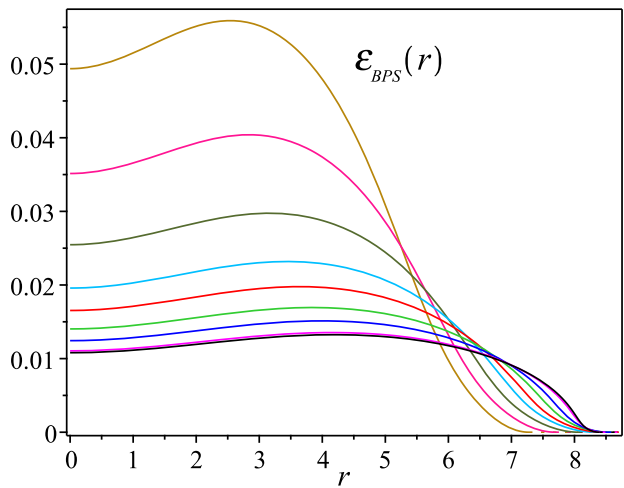}\vspace{-0.25cm} }
\caption{The BPS energy density $\protect\varepsilon_{_{BPS}}(r)$.}
\label{fig4}
\end{figure}

In our first analysis we have solved the BPS equations (\ref{bps1h}) and (%
\ref{bps2h}) by fixing $N=1$, $\kappa=1$, $W_{0}=1$ and $\lambda=2.5$ and
running the electromagnetic coupling constant $g$. The compacton solutions
are depicted in Figs. \ref{fig1}, \ref{fig2}, \ref{fig3} and \ref{fig4}.

\begin{figure}[H]
\centering\includegraphics[height=6.5cm]{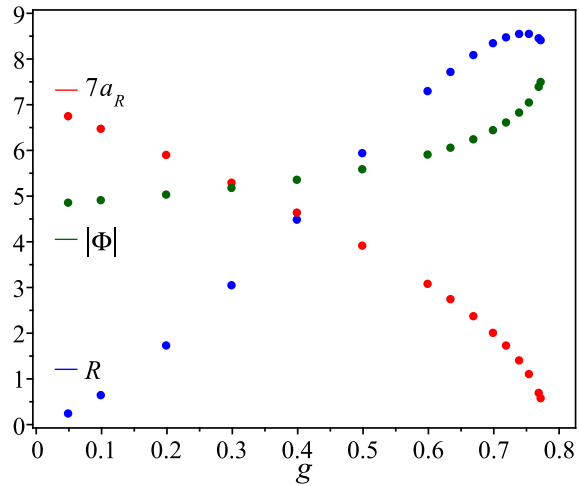}\vspace{-0.25cm}
\caption{Behaviors of the compacton radius $R\;vs.\;g$ (blue color), the
gauge vacuum value $a_{_{R}}\;vs.\;g$ (red color) and the magnetic flux $%
|\Phi|\;vs.\;g$ (green color) of compacton solutions of the BPS equations (%
\protect\ref{bps1h}) and (\protect\ref{bps2h}).}
\label{fig5}
\end{figure}

The Skyrme field profile $h(r)$ is depicted in Fig. \ref{fig1} for various
values of $g$. The colored solid lines represents the profiles of the $h(r)$
in the interval $0\leq r\leq R$ and the respective colored pointed lines
represent the vacuum value, $h=1$, in the interval $R\leq r<\infty$. The
compacton radius $R$ for various values of $g$ are shown in Fig. \ref{fig5}.

The Figure \ref{fig2} depicts the gauge field profile $a(r)$. Similar to the
description given in Fig. \ref{fig1}, the colored solid lines represents the
profiles of the $a(r)$ in the interval $0\leq r\leq R$ and the respective
colored pointed lines represent the vacuum value, $a(R)=a_{_{R}}$, in the
interval $R\leq r<\infty$. The profiles show the vacuum value, $%
a(R)=a_{_{R}} $, diminishes whenever $g$ increases.

The profiles of the magnetic field are presented in Figs. \ref{fig3}. They
are ring structures whose maximum for small $g$ is located close to the
origin whereas for sufficiently larger values of $g$ the maximum moves its
position to very near of the frontier of the compacton. The amplitude of the
maximum value of the magnetic field is greater for small values of $g$,
i.e., in our case it means $0<g<0.5$, it is not show in the figure but it
can be seen clearly from the behavior given by Eq. (\ref{magr=0}).

The profiles of the BPS energy density (\ref{edbps}) are presented in Figs. %
\ref{fig4}. The behavior at origin is given by (see Eq. (\ref{ebpsr=0})),
\begin{eqnarray}
\varepsilon_{_{BPS}} &=& \frac{23.04\times10^{-4}}{g^{6}}  \notag \\%
[0.2cm]
& & +\frac{5898.24\times10^{-8}}{g^{9}}\left[ g-0.1875\right] r^{2}+....
\end{eqnarray}
For small values of $g$ the profiles look like lumps centered at origin. In
our case it happens for $0<g<0.1875$, the profiles are not presented in Fig. %
\ref{fig4} because their amplitudes are bigger than the ones shown there.
For sufficiently large values of $g$ the profiles acquire a ring-like form,
in our case, the Fig. \ref{fig4} shows such structures for $g>0.6$.

The dependence of the compacton radius $R\;vs.\;g$, the gauge vacuum value $%
a_{_{R}}\;vs.\;g$ and the total magnetic flux $|\Phi|\;vs.\;g$ (by fixing
all other parameters) are shown in Fig. \ref{fig5} for $N=1$, $\kappa=1$, $%
W_{0}=1$ and $\lambda=2.5$. We have observed the vacuum value, $%
a(R)=a_{_{R}} $, diminishes whenever $g$ increases, i.e., $%
a_{_{R}}\rightarrow0$ for sufficiently large values of $g$. Consequently, we
get
\begin{equation}
\Phi\rightarrow-\frac{2\pi}{g}N, \quad \mathcal{Q}_{\text{em}}\rightarrow%
\frac{2\pi\kappa}{g^{2}}N,  \label{fluxN}
\end{equation}
being, therefore, quantities topologically quantized.

From the Figs. \ref{fig2} and \ref{fig5}, we observe clearly by considering
the electromagnetic coupling constant $g$ the only free parameter (and all
other fixed) the compacton radius $R$ possess a maximum value.

\begin{figure}[H]
\centering{\includegraphics[height=6.5cm]{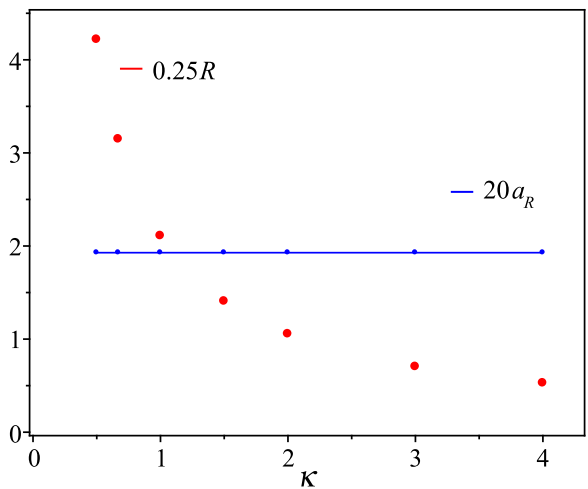}\vspace{-0.25cm} }
\caption{Compacton radius $R\;vs.\;\protect\kappa$ (red dots) and the gauge
field vacuum value $a_{_{R}}\;vs.\;\protect\kappa$ (blue dotted-line).}
\label{fig6}
\end{figure}

Similarly, we have analyzed the dependence of the compacton radius $%
R\;vs.\;\kappa$ and the gauge vacuum value $a_{_{R}}\;vs.\;\kappa$ (by
fixing all other parameters). The numerical analysis have shown that the
radius is inversely proportional to $\kappa$ ($R \propto\kappa^{-1}$)
whereas the gauge vacuum value $a_{_{R}}$ remains constant. The Fig. \ref%
{fig6} shows $R$ vs. $\kappa$ for $N=1$, $g=0.77$, $W_{0}=1$ and $%
\lambda=2.5 $ whereas $a_{_{R}}=0.0964097741$ for all values of $\kappa$.

\begin{figure}[H]
\centering{\includegraphics[height=6.5cm]{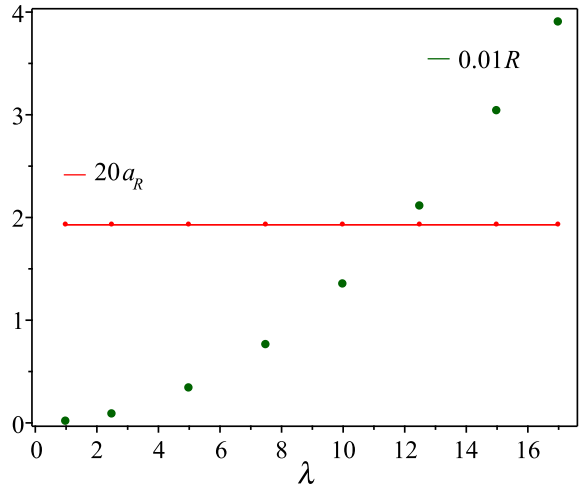}\vspace{-0.25cm} }
\caption{Compacton radius $R\;vs.\;\protect\lambda$ (green dots) and the
gauge field vacuum value $a_{_{R}}\;vs.\;\protect\lambda$ (red dotted-line).}
\label{fig7}
\end{figure}

Our third analysis has looked the dependence of compacton radius $%
R\;vs.\;\lambda$ and the gauge vacuum value $a_{_{R}}\;vs.\;\lambda$ (for
all others parameters set). The numerical analysis have shown that the
radius depends quadratically with $\lambda$ ($R \propto\lambda^{2}$) whereas
the gauge vacuum value $a_{_{R}}$ remains constant. Such a dependence is
depicted in Fig. \ref{fig7} for $N=1$, $g=0.77$, $W_{0}=1$ and $\kappa=1$
whereas $a_{_{R}}=0.0964097741$ for all values of $\lambda$.

Until now, our analysis of the compacton solitons allows to conclude the
gauge vacuum value $a_{_{R}}$ only depends in the electromagnetic coupling
constant $g$, consequently, the total magnetic flux is independent from the
values of $\kappa$ and $\lambda$ and it becomes quantized for sufficiently
large values of $g$.

\subsection{Solitons with exponential-law decay}

We have solved the BPS equations (\ref{bps1h}) and (\ref{bps2h}) for the
following superpotential
\begin{equation}
W(h)=W_{0}(1-h)^{2},  \label{exp_decay}
\end{equation}
which provides the potential
\begin{equation}
\mathcal{V}(h)=\frac{\kappa^{4}W_{0}^{2}}{2\lambda^{6}g^{6}}(1-h)^{2}.
\end{equation}
Similar potential was used in \cite{adam2}.

\begin{figure}[H]
\centering{\includegraphics[height=6.5cm]{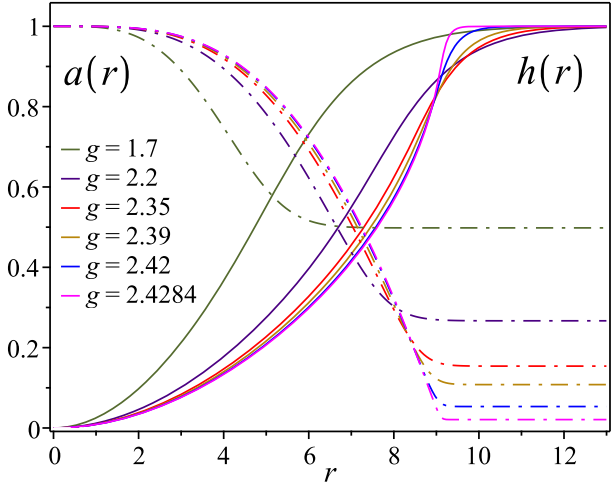}\vspace{-0.25cm} }
\caption{Profiles of the Skyrme field $h(r)$ and the gauge field $a(r)$ with
exponential decay engendered by the superpotential (\protect\ref{exp_decay}). }
\label{fig8}
\end{figure}
\begin{figure}[H]
\centering{\includegraphics[height=6.5cm]{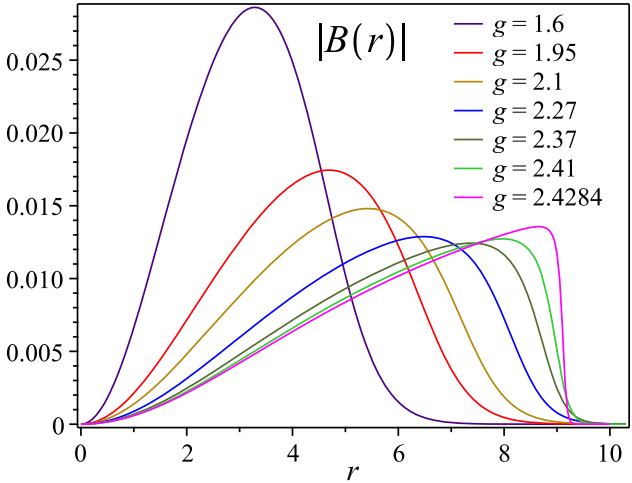}\vspace{-0.25cm} }
\caption{Profiles of the magnetic field $B(r)$ with exponential decay
engendered by the superpotential (\protect\ref{exp_decay}).}
\label{fig9}
\end{figure}
\begin{figure}[ ]
\centering{\includegraphics[height=6.5cm]{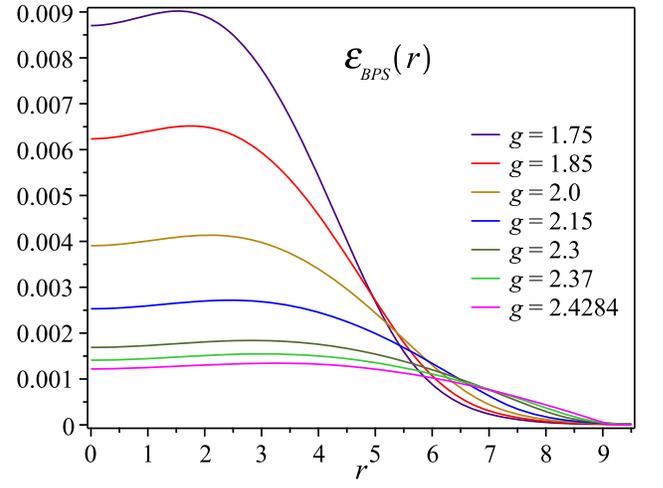}\vspace{-0.25cm} }
\caption{Profiles of BPS energy density $\protect\varepsilon_{_{BPS}}(r)$
with exponential decay engendered by the superpotential (\protect\ref%
{exp_decay}).}
\label{fig10}
\end{figure}

We have performed our analysis by solving the BPS equations (\ref{bps1h})
and (\ref{bps2h}) by setting $N=1$, $\kappa=1$, $W_{0}=0.5$, $\lambda=1$ and
various values of $g$. The Figs. \ref{fig8}, \ref{fig9} and \ref{fig10}
present the profiles of the Skyrme and gauge field, the magnetic field and
the BPS energy, respectively, for increasing values of $g$. We observe, for
sufficiently large value of $g$, the profiles acquire a compactonlike
structure.

\begin{figure}[ ]
\centering{\includegraphics[height=6.5cm]{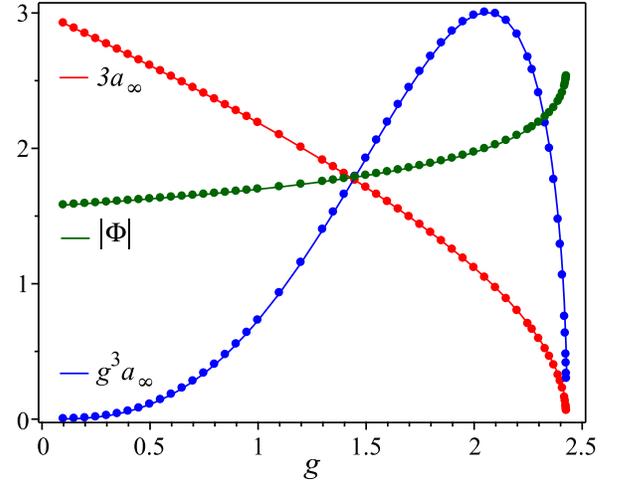}\vspace{-0.25cm} }
\caption{Behaviors of the gauge vacuum value $a_{\infty}\;vs.\;g$ (red
color), the magnetic flux $|\Phi|\;vs.\;g$ (green color) and the quantity $%
g^{3}a_{\infty}\;vs.\;g$ (blue color) of the solutions with exponential
decay engendered by the superpotential (\protect\ref{exp_decay}).}
\label{fig11}
\end{figure}

The behaviors of the gauge vacuum value $a_{\infty}\;vs.\;g$ (red color) and
the magnetic flux $|\Phi|\;vs.\;g$ (green color) depicted in Fig. \ref{fig11}
show a similar structure with the one observed in compacton case. Similarly,
we show the behavior of the quantity $g^{3}a_{\infty}\;vs.\;g$ (blue color)
which controls the spread (\ref{massa2}) of the solutions for sufficiently
large values of $r$.

The numerical analysis shows the gauge vacuum value $a_{_{\infty}}$ only
depends on the values of the electromagnetic coupling constant $g$ such it
happens in the compacton case.

\subsection{Solitons with power-law decay}

To obtain BPS solitons with power-law decay from solving the BPS equations (%
\ref{bps1h}) and (\ref{bps2h}), we consider the following superpotential
\begin{equation}
W(h)=W_{0}(1-h)^{\sigma},  \label{power_law}
\end{equation}
with $\sigma>2$, providing the potential
\begin{equation}
\mathcal{V}(h)=\frac{\kappa^{4}W_{0}^{2}\sigma^{2}}{8\lambda^{6}g^{6}}%
(1-h)^{2(\sigma-1)}.
\end{equation}

\begin{figure}[ptb]
\centering{\includegraphics[height=6.5cm]{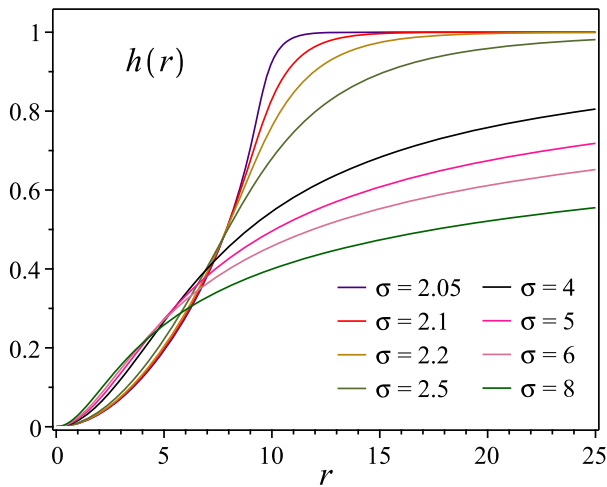}\vspace{-0.25cm} }
\caption{The profiles $h(r)$ of the Skyrme field with power-law decay
generated by the superpotential (\protect\ref{power_law}).}
\label{fig12}
\end{figure}

\begin{figure}[ptb]
\centering{\includegraphics[height=6.5cm]{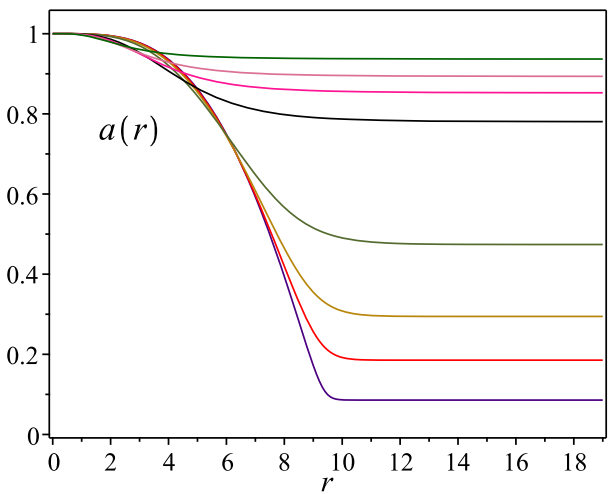}\vspace{-0.25cm} }
\caption{The gauge field profiles $a(r)$ with power-law decay generated by
the superpotential (\protect\ref{power_law}).}
\label{fig13}
\end{figure}

\begin{figure}[ptb]
\centering{\includegraphics[height=6.5cm]{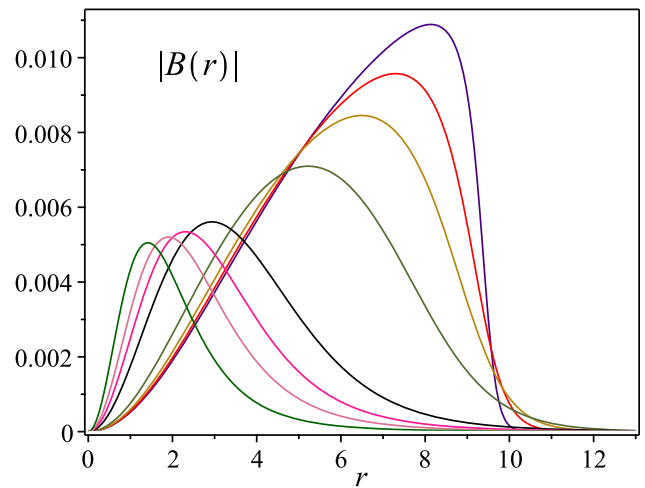}\vspace{-0.25cm} }
\caption{The magnetic field profiles $B(r)$ with power-law decay generated
by the superpotential (\protect\ref{power_law}).}
\label{fig14}
\end{figure}

\begin{figure}[ptb]
\centering{\includegraphics[height=6.5cm]{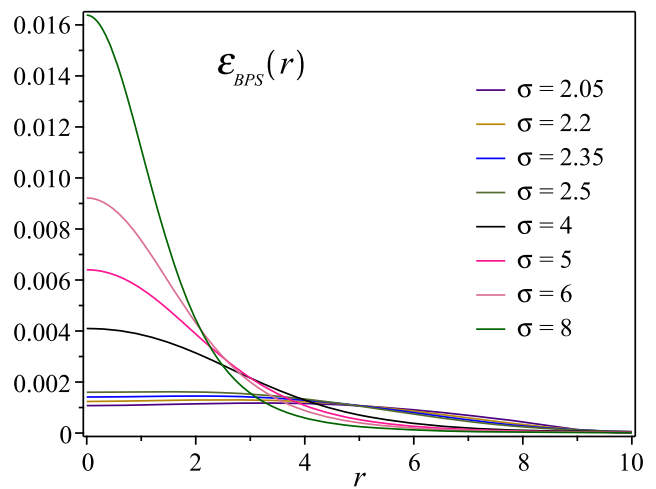}\vspace{-0.25cm} }
\caption{The BPS energy density $\protect\varepsilon_{_{BPS}}(r)$ with
power-law decay generated by the superpotential (\protect\ref{power_law}). }
\label{fig15}
\end{figure}

We have performed our analysis by solving the BPS equations (\ref{bps1h})
and (\ref{bps2h}) by setting $N=1$, $\kappa=1$, $W_{0}=0.5$, $\lambda=1$, $%
g=2.5$ and various values of the parameter $\sigma$. The Figs. \ref{fig12}, %
\ref{fig13}, \ref{fig14} and \ref{fig15} depict the profiles of the Skyrme
field $h(r)$, the gauge field $a(r)$, the magnetic field $B(r)$ and the BPS
energy density $\varepsilon_{_{BPS}}(r)$, respectively. The behavior of the
profiles becomes similar to a compactonlike form when the parameter $%
\sigma\rightarrow2$.

Similarly, like it happens in the two previous case the numerical analysis
again shown, for fixed value of $\sigma>2$, the gauge vacuum value $%
a_{_{\infty}}$ only depends of the values of the electromagnetic coupling
constant $g$. From the Fig. \ref{fig13}, for a fixed value of $g$, we see
the value of $a_{_{\infty}} \rightarrow1$ for $\sigma\gg2$ implying the
total magnetic flux $\Phi\rightarrow0$.

\section{Remarks and conclusions}

We have shown the existence of BPS solitons in a type of generalized Chern-Simons baby Skyrme model (\ref{lag1a}) where the generalized function  $W(\phi_n)$ coupled to the sigma-model term becomes the superpotential which defines the self-dual potential (\ref{condition}). The guidelines for the construction of such a BPS model (\ref{lag1}) or (\ref{lag1a}) are provided by the unsuccessful implementation of the BPS formalism in the Chern-Simons restricted baby Skyrme model introduced in Eq. (\ref{lag0}). The successful implementation of the BPS formalism in the model (\ref{lag1}) has allowed to obtain an energy lower-bound (BPS limit) and the self-dual equations satisfied by the field saturating such a limit. The BPS energy is proportional to the topological charge of the Skyrme field so it is quantized. On the other hand, the total magnetic flux and total electric charge are proportional to each other but in general are not quantized. However, for sufficiently large values of the electromagnetic coupling constant $g$ both become quantized (see Eq. (\ref{fluxN})).

The superpotential plays the principal role defining the BPS solitons thus we have considered it being a well-behaved function in the whole target space. We have observed the existence of three classes of self-dual solutions closely related with the behavior of the superpotential. The first class of solitons we have obtained are the so-called compactons, which arise when the superpotential behaves like $W(r) \approx(1-h(r))^{\sigma}$ for $r\rightarrow R$ and $1<\sigma<2$, where $R$ is the compacton radius. The other two classes of solitons are noncompacton structures, i.e., they are regular functions in $0\leq r<\infty$ but they are different because their respective tails have different behaviors for $r\rightarrow \infty$. Thus, the first noncompacton solitons are generated by superpotential behaving like $W(r)\approx(1-h(r))^2$ for $r\rightarrow \infty $ whose tail decays following an exponential-law $e^{-\alpha r^{2}}$ ($\alpha>0$). The second class of noncompacton solitons possess a tail following a power-law decay $r^{-\beta}$ ($\beta>0$) for $r\rightarrow \infty $ and the superpotential behaves like $W(r)\approx (1-h(r))^{\sigma}$ with $\sigma>2$. Depending of the parameter values the two last solitons can exhibit a compactonlike behavior.

We are investigating the existence of BPS solitons in a baby Skyrme model
gauged with the Maxwell-Chern-Simons action and into the presence of Lorentz
violation.

\begin{acknowledgments}
This study was financed in part by the Coordena\c{c}\~ao de Aperfei\c{c}%
oamento de Pessoal de N\'{\i}vel Superior - Brasil (CAPES) - Finance Code
001. We thank also the Conselho Nacional de Desenvolvimento Cient{\'\i}fico
e Tecnol\'ogico (CNPq), and the Funda\c{c}\~ao de Amparo \`a Pesquisa e ao
Desenvolvimento Cient{\'\i}fico e Tecnol\'ogico do Maranh\~ao (FAPEMA)
(Brazilian Government agencies). In particular, ACS, CFF and ALM thank the
full support from CAPES and RC acknowledges the support from the grants
CNPq/306385/2015-5 and FAPEMA/Universal-01131/17.
\end{acknowledgments}

\end{document}